\documentclass{cernyrep}
\usepackage{texnames}
\usepackage[T1]{fontenc}
\usepackage{units}

\usepackage{varwidth}
\usepackage{xcolor}

\title{An Overview of Recent Progress in Laser Wakefield Acceleration Experiments}
\author{S.P.D. Mangles}
\institute{John Adams Institute for Accelerator Science, Blackett Laboratory,
Imperial College London, UK}
\date{}                         

\begin{document}

\maketitle
\begin{abstract}
The goal of this paper is to examine experimental progress in laser wakefield acceleration over the past decade (2004--2014), and to use trends in the data to understand some of the important physical processes. By examining a set of over 50 experiments, various trends concerning the relationship between plasma density, accelerator length, laser power and the final electron beam energy are revealed. The data suggest that current experiments are limited by dephasing and that current experiments typically require some pulse evolution to reach the trapping threshold.
\end{abstract}

\section{Introduction}
This paper is a summary of a lecture given at the CERN Accelerator School 2015 on plasma-based wakefield accelerators.
Its purpose is to provide an overview of recent experimental progress in laser wakefield acceleration, concentrating on the energy frontier and general trends that can be observed in the data produced by the many groups around the world contributing to this growing field.
There are now more than 20 active laboratories performing laser wakefield acceleration ex\-peri\-ments.
We will not detail the key results from each of these laboratories in this paper, although we will examine data from various published experiments and use the trends in the data to try to gain some understanding of  the underlying physical processes.

Laser wakefield accelerators were proposed 36 years ago in the seminal 1979 paper by Tajima and Dawson \cite{Tajima1979} and experiments in laser wakefield acceleration were  undertaken as soon as laser pulses of sufficiently short duration and high power became available, thanks to the development of the laser technique called `chirped pulse amplification' \cite{Strickland1985}.
Some of the early work in what we call laser wakefield acceleration (\ie where the pulse duration, $\tau_\mathrm{L}$ is comparable to the plasma period $2\pi / \omega_\mathrm{p}$) occurred in the 1990s (\eg
Refs.\cite{Nakajima1994, Amiranoff1998}) using picosecond glass lasers.
  The field passed a major milestone in the early 2000s as high-power (${\sim}10\UW[T] $) femtosecond laser pulses, using titanium sapphire laser systems, became available.
 One major result from that era occurred when three groups from the UK, USA and France all demonstrated that laser wakefield accelerators could produce  electron beams with well-defined energies \cite{Mangles2004, Geddes2004,Faure2004}.
 The electron beams produced by these ${\sim}10\UW[T]$ lasers had energies of ${\sim}100\UMeV$ and were produced in plasmas of a plasma density  ${\sim}10^{19}\Ucm^{-3}$ that were only ${\sim}1\Umm$ long.

One of the key challenges that has driven progress in the field of laser wakefield acceleration is the maximum achievable beam energy,  and this paper will concentrate on this challenge.
There have, of course, been many other significant areas of experimental progress including: improving beam stabil\-ity, especially by controlling injection (\eg Refs.\cite{Faure2006,Gonsalves2011}); diagnosis of  wakefield accelerators (\eg Refs.\cite{Matlis2006,Savert2015}) and the resulting improvements in our understanding  of the underlying processes; and the application of laser wakefield accelerators for a range of applications, perhaps most notably their use as novel sources of X-radiation (\eg Refs.\cite{Rousse2004,Kneip2010,Fuchs2009b}).
Details of the progress in these areas are outside the scope of this paper.

\begin{figure}[h!]
\begin{center}
\includegraphics[width=10cm]{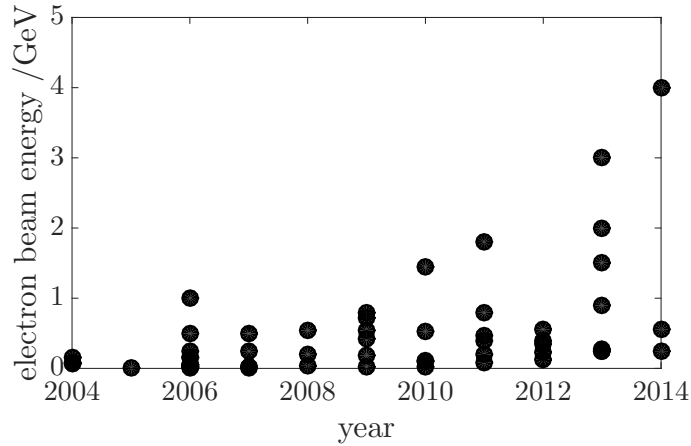}
\caption{Reported electron beam energies from laser wakefield experiments at various laboratories over the last decade; data from Refs.
\cite{ Mangles2004,
 Geddes2004,
 Faure2004,
 Miura2005,
 Kotaki2006,
 Mori2006,
 Masuda2006,
 Hsieh2006,
 Hidding2006,
 Hosokai2006a,
 Mangles2006a,
 Faure2006,
 Reed2006,
 Leemans2006,
 Masuda2007,
 Ohkubo2007,
 Karsch2007,
 Mangles2007,
 Gamucci2008,
 Rowlands-Rees2008,
 Hafz2008a,
 Schmid2009a,
 Kneip2009b,
 Froula2009,
 Schmid2010a,
 Pak2010a,
 Ibbotson2010a,
 Clayton2010,
 McGuffey2010,
 Lu2011,
 Fourmaux2011a,
 Liu2011a,
 Pollock2011,
 Lundh2011,
 Gonsalves2011,
 Brijesh2012,
 Mo2012,
 Kneip2012,
 Weingartner2012,
 Albert2013,
 Walker2013,
 Mo2013a,
 Corde2013b,
 Chen2013a,
 Kim2013,
 Wang2013a,
 Sarri2014,
 Powers2014,
 Leemans2014,
 Khrennikov2015,
 Schnell2015
}.
}
\label{energy-year}
\end{center}
\end{figure}

This paper will examine how the maximum achievable beam energy has progressed over the last decade, and will use a set of published results
\cite{Mangles2004,
 Geddes2004,
 Faure2004,
 Miura2005,
 Kotaki2006,
 Mori2006,
 Masuda2006,
 Hsieh2006,
 Hidding2006,
 Hosokai2006a,
 Mangles2006a,
 Faure2006,
 Reed2006,
 Leemans2006,
 Masuda2007,
 Ohkubo2007,
 Karsch2007,
 Mangles2007,
 Gamucci2008,
 Rowlands-Rees2008,
 Hafz2008a,
 Schmid2009a,
 Kneip2009b,
 Froula2009,
 Schmid2010a,
 Pak2010a,
 Ibbotson2010a,
 Clayton2010,
 McGuffey2010,
 Lu2011,
 Fourmaux2011a,
 Liu2011a,
 Pollock2011,
 Lundh2011,
 Gonsalves2011,
 Brijesh2012,
 Mo2012,
 Kneip2012,
 Weingartner2012,
 Albert2013,
 Walker2013,
 Mo2013a,
 Corde2013b,
 Chen2013a,
 Kim2013,
 Wang2013a,
 Sarri2014,
 Powers2014,
 Leemans2014,
 Khrennikov2015,
 Schnell2015
}
to try to understand some of the physics behind this progress.
The progress in the maximum beam energy in laser wakefield accelerator ex\-peri\-ments has been rapid, as shown in \Fref{energy-year}, from a maximum beam energy of $0.2\UGeV$ reported in 2002 \cite{Malka2002} to the current record of $4\UGeV$ from the group at the Lawrence Berkeley National Laboratory \cite{Leemans2014} achieved in 2014 -- an increase by a factor of 20 in just over a decade. It should be noted that this is by no means an exhaustive list of all published experiments in laser wakefield accelerators (there are just 52 publications in this dataset, whereas a literature search for papers on `laser wakefield' will find over 1000 papers).

\section{Overall trends in laser wakefield acceleration experiments }

\begin{figure}
\begin{center}
\includegraphics[width=10cm]{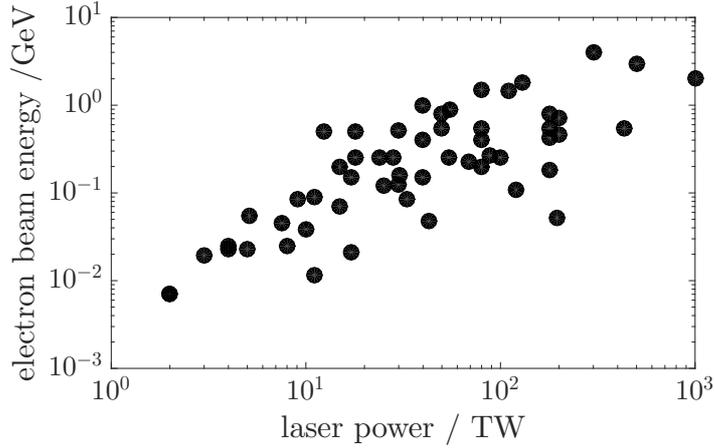}
\caption{Variation of reported electron beam energy with laser power from various experiments; data from Refs.
\cite{ Mangles2004,
 Geddes2004,
 Faure2004,
 Miura2005,
 Kotaki2006,
 Mori2006,
 Masuda2006,
 Hsieh2006,
 Hidding2006,
 Hosokai2006a,
 Mangles2006a,
 Faure2006,
 Reed2006,
 Leemans2006,
 Masuda2007,
 Ohkubo2007,
 Karsch2007,
 Mangles2007,
 Gamucci2008,
 Rowlands-Rees2008,
 Hafz2008a,
 Schmid2009a,
 Kneip2009b,
 Froula2009,
 Schmid2010a,
 Pak2010a,
 Ibbotson2010a,
 Clayton2010,
 McGuffey2010,
 Lu2011,
 Fourmaux2011a,
 Liu2011a,
 Pollock2011,
 Lundh2011,
 Gonsalves2011,
 Brijesh2012,
 Mo2012,
 Kneip2012,
 Weingartner2012,
 Albert2013,
 Walker2013,
 Mo2013a,
 Corde2013b,
 Chen2013a,
 Kim2013,
 Wang2013a,
 Sarri2014,
 Powers2014,
 Leemans2014,
 Khrennikov2015,
 Schnell2015
}.
 }
\label{energy-power}
\end{center}
\end{figure}

The rapid progress shown in \Fref{energy-year} is impressive.
But how has it been achieved?
Over the same period of time, short-pulse (${\approx}30\Ufs$) laser systems have become more powerful.
Figure \ref{energy-power} shows that there is a  clear trend: higher-power lasers are capable of producing higher-energy electron beams.
However,  these gains were not achieved by simply increasing the laser power, the researchers behind these experiments have often found the optimum conditions for their experiments.
Key parameters involved in this optimization include the operating plasma density, the length of the accelerator and the laser intensity.
This section will examine the data from various experiments and compare them with  predicted trends, to see whether they can confirm those predictions and the underlying physical processes.

\subsection{Accelerator length and operating plasma density}
Let us first consider the density of the plasma accelerator.
The energy gained by a  particle of charge $q$ in an accelerating structure is simply  proportional to the product of the electric field and the length of the accelerator, $d$, \ie $W  \simeq qEd$, where $E$ is the average  accelerating electric field experienced by the particle.
One of the key physical limitations in a laser wakefield accelerator is dephasing.
The electrons trapped in a wake are highly relativistic ($\gamma \gg 1$), so they travel at a speed approaching that of light in vacuum ($v_\mathrm{e} \to c$), but the phase speed of the wake is determined by the speed of the laser pulse that drives the plasma wave.
A simple expression for the speed of a laser pulse in a plasma can be found by using the standard expression for the group velocity of an electromagnetic wave in a plasma,
\begin{equation}
\frac{v_\mathrm{g}}{c} =  \sqrt{ 1 - \frac{n_\mathrm{e}}{n_\mathrm{c}}} \approx  1 - \frac{1}{2} \frac{n_\mathrm{e}}{n_\mathrm{c}}\ ,
\end{equation}
where $n_\mathrm{e}$ is the electron density of the plasma, $n_\mathrm{c}$ is the critical density for propagation of the electro\-magnetic (\ie when the plasma frequency $\omega_\mathrm{p}$ equals the frequency of the electromagnetic wave, $\omega_0$) and it is assumed that $n_\mathrm{e} \ll n_\mathrm{c}$.
The wake's phase speed is therefore slightly, but significantly, less than $c$; crucially, the lower the plasma density, the faster the phase velocity.


Because of this difference between the electron velocity and the wake phase velocity,  electrons in a laser-driven wake will outrun the wake\footnote{Note that  dephasing does not occur in wakefield accelerators driven by highly relativistic charged particle beams as both the accelerated and driver beams are highly relativistic.}.
If the electron is injected at the start of the accelerating phase of the plasma wave and then outruns the wave by half a plasma wavelength (\ie $\lambda_\mathrm{p}/2  = \pi c/\omega_\mathrm{p}$), it can no longer gain energy from the plasma wave.
If the electron has an initial velocity $v_\mathrm{e} = \beta_\mathrm{e} c$ and the plasma wave has a phase velocity $v_\phi = \beta_\phi c$, then the time it takes for this to occur is $t_\mathrm{d} = \lambda_\mathrm{p}/ (2c(\beta_\mathrm{e} - \beta_\phi ) )$.
The dephasing length is then the distance that the electron travels in this time.
Since $\beta_\mathrm{e} \to 1$ and $\beta_\phi \simeq  1 - \frac{1}{2}  ({n_\mathrm{e}}/{n_\mathrm{c}})$, this reduces to
\begin{equation}
\label{dephasing_length}
  L_{\rm dephasing}  \simeq \frac{n_\mathrm{c}}{n_\mathrm{e}} \lambda_\mathrm{p}   \propto n_\mathrm{e}^{-\frac{3}{2}}\ .
\end{equation}

\begin{figure}
\begin{center}
\includegraphics[width=10cm]{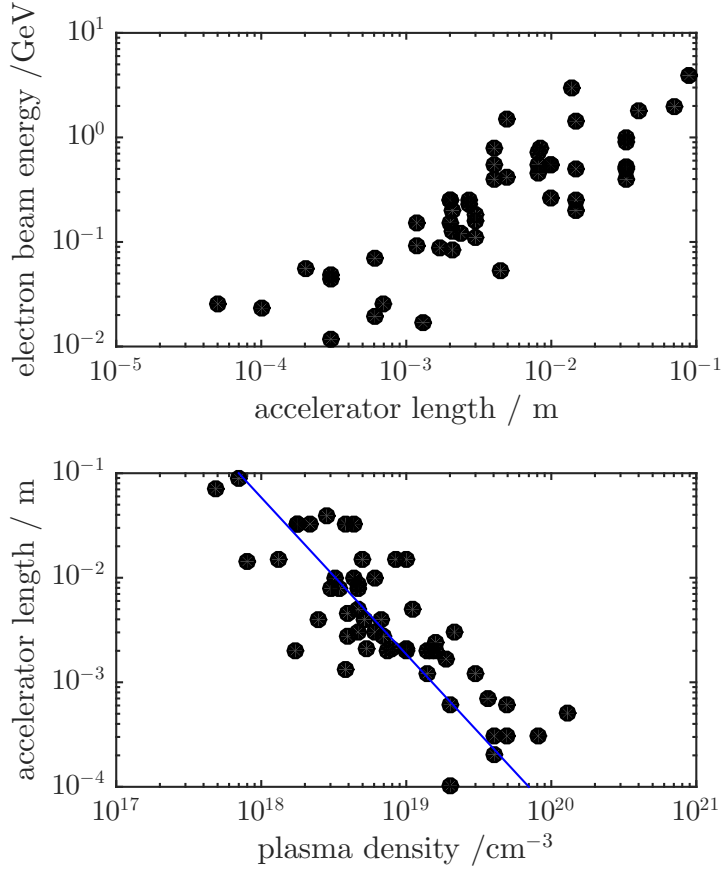}
\caption{Top: Variation of  reported electron beam energy with accelerator length.
Bottom:  Relationship between operating plasma density and accelerator length.  The line shows the expression for the dephasing length (\Eref{dephasing_length}).
Data from Refs.
\cite{Mangles2004,
 Geddes2004,
 Faure2004,
 Miura2005,
 Kotaki2006,
 Mori2006,
 Masuda2006,
 Hsieh2006,
 Hidding2006,
 Hosokai2006a,
 Mangles2006a,
 Faure2006,
 Reed2006,
 Leemans2006,
 Masuda2007,
 Ohkubo2007,
 Karsch2007,
 Mangles2007,
 Gamucci2008,
 Rowlands-Rees2008,
 Hafz2008a,
 Schmid2009a,
 Kneip2009b,
 Froula2009,
 Schmid2010a,
 Pak2010a,
 Ibbotson2010a,
 Clayton2010,
 McGuffey2010,
 Lu2011,
 Fourmaux2011a,
 Liu2011a,
 Pollock2011,
 Lundh2011,
 Gonsalves2011,
 Brijesh2012,
 Mo2012,
 Kneip2012,
 Weingartner2012,
 Albert2013,
 Walker2013,
 Mo2013a,
 Corde2013b,
 Chen2013a,
 Kim2013,
 Wang2013a,
 Sarri2014,
 Powers2014,
 Leemans2014,
 Khrennikov2015,
 Schnell2015
}.}
\label{density-length}
\end{center}
\end{figure}

It is interesting to see how the lengths of the accelerators  in the set of experiments vary, and how this compares with what we might expect if dephasing is important.
These data are shown in \Fref{density-length}.
The top panel shows how the reported electron beam energy varies with the length of the wakefield acceler\-ator.  There is a clear correlation -- the higher electron energies are achieved with longer accelerators, as we might expect.
The bottom panel of \Fref{density-length} shows how the length of the accelerator and the plasma density at which it was operating are related.
The line on this curve is the simple expression for the dephasing length (\Eref{dephasing_length}).

The maximum electric field that a plasma wave can support increases with plasma density, since it scales as

\begin{equation}
E_{\mathrm {max}} \simeq m_\mathrm{ c} \omega_\mathrm{p} / e \propto \sqrt{n_\mathrm{e}}\ .
\end{equation}

\begin{figure}
\begin{center}
\includegraphics[width=10cm]{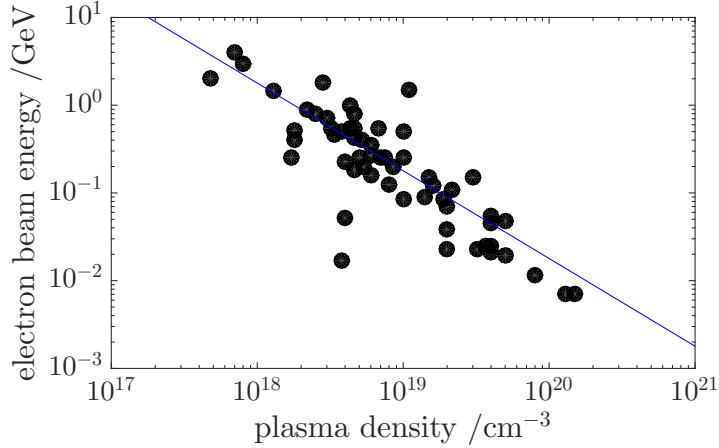}
\caption{Variation of reported electron beam energy with the density in the accelerator.
The line shows the relation $W/(m_\mathrm{e}c^2) =  \kappa\, n_\mathrm{c} / n_\mathrm{e}$ with $\kappa = 1$.
Data from Refs.
\cite{Mangles2004,
 Geddes2004,
 Faure2004,
 Miura2005,
 Kotaki2006,
 Mori2006,
 Masuda2006,
 Hsieh2006,
 Hidding2006,
 Hosokai2006a,
 Mangles2006a,
 Faure2006,
 Reed2006,
 Leemans2006,
 Masuda2007,
 Ohkubo2007,
 Karsch2007,
 Mangles2007,
 Gamucci2008,
 Rowlands-Rees2008,
 Hafz2008a,
 Schmid2009a,
 Kneip2009b,
 Froula2009,
 Schmid2010a,
 Pak2010a,
 Ibbotson2010a,
 Clayton2010,
 McGuffey2010,
 Lu2011,
 Fourmaux2011a,
 Liu2011a,
 Pollock2011,
 Lundh2011,
 Gonsalves2011,
 Brijesh2012,
 Mo2012,
 Kneip2012,
 Weingartner2012,
 Albert2013,
 Walker2013,
 Mo2013a,
 Corde2013b,
 Chen2013a,
 Kim2013,
 Wang2013a,
 Sarri2014,
 Powers2014,
 Leemans2014,
 Khrennikov2015,
 Schnell2015
}.}
\label{energy-density}
\end{center}
\end{figure}

The maximum energy that can be gained by an electron in a plasma wave as a function of plasma density is therefore expected to be
\begin{equation}
 W(n_\mathrm{e}) \simeq E_{\max} L_{\mathrm{dephasing}}    \propto  \frac{1}{n_\mathrm{e}}\ .
\end{equation}
Figure \ref{energy-density} shows how the  plasma density and beam energy vary in the set of experiments.
The line on \Fref{energy-density} is simply $W/(m_\mathrm{e}c^2) =  \kappa\, n_\mathrm{c}/n_\mathrm{e}$  and shows good agreement with the entire dataset for $\kappa = 1$.
The scaling laws in Ref.\cite{Lu2007e}, by Wei Lu \textit{et al.}, for the blow-out or `bubble' regime of wakefield accelerators, suggest that the scaling law should be
\begin{equation}
\label{}
 W(n_\mathrm{e}, a_0) \simeq  \frac{2}{3} a_0 \frac{n_\mathrm{c}}{n_\mathrm{e}} m_\mathrm{e} c^2  \propto \frac{a_0}{n_\mathrm{e}}\ ,
\end{equation}
where $a_0  = eA_0/ (m_\mathrm{e} c) = e E_0 / (m_\mathrm{e}\omega_0 c)$ is the normalized peak vector potential (or strength parameter) of the laser pulse.
This scaling predicts that the beam energy should not only be proportional to $1/n_\mathrm{e}$ but also proportional to the laser strength, $a_0$. The experiments shown correspond to a wide range of initial laser intensities (corresponding to  $a_0 = 0.5$--$4.0$), yet they do not appear to show a dependence on $a_0$.

One possible reason for this apparent discrepancy is that the initial value for $a_0$ is not the value of $a_0$ that determines the wake amplitude.
It is well known that  laser pulses can undergo significant evolution once they enter the plasma. The processes of self-focusing, self-compression and photon acceleration\cite{Mori1997a} can all act to change $a_0$ as the pulse propagates. Together, these processes can be termed the `self-evolution' of the laser pulse.
  One interpretation of the experimental data is, therefore, that the process of self-evolution has occurred until $a_0 \approx 3$ for all of the data shown.
  Why would this be the case?
  One reasonable hypothesis is that each  point in the dataset corresponds to the maximum energy achieved  during a particular experiment and that this will occur at (or at least close to) the lowest density at which a particular experiment can trap and accelerate electrons.
Many experiments  operate by fixing the laser power and plasma length while varying the plasma density, for reasons of experimental simplicity.
When an experiment is conducted in this manner, there will be a minimum density at which electron beams are trapped and accelerated (the trapping threshold).
  Because self-evolution happens less quickly and severely at lower densities, the maximum $a_0$ that is reached inside the accelerator will decrease with decreasing plasma density.
  Therefore, the maximum achieved electron beam energy will correspond to the minimum laser strength required to produce trapping.  The fact that the experimental dataset matches the non-linear wakefield scaling but only if $a_0 \approx 3$ suggests that the minimum $a_0$ required for trapping is $a_0 \approx 3$.


\subsection{Laser spot size and matched guiding}

The  experimental data  clearly show that higher-power lasers are required to achieve higher electron beam energies.
But what are the physical processes behind this trend?
It was argued that the experimental trends are consistent with there being a minimum value  of $a_0 \approx 3$, which is needed for trapping, and that this value is reached because of the way the pulse evolves as it propagates.

Consider the relationship between the intensity, $I$ ($\propto a_0^2$), and power, $P$, of a laser pulse.
Since  $I = P/A$, where $A$ ($\propto w^2$,  the laser spot size) is the focal spot area, we have

\begin{equation}
\label{}
    P \propto a_0^2 w^2  \,.
\end{equation}
The fact that higher-power lasers are needed to reach the $a_0 \approx 3$ threshold at lower densities therefore implies that the spot size that these laser pulses produce after evolution is larger.
Pulse evolution is a result of the feedback between the refractive index gradient associated with the plasma wave and the laser pulse, \ie it is mediated by the plasma itself.
Lower-density plasmas, therefore, have a lesser effect on the laser pulse -- resulting in slower evolution -- and crucially this affects the properties that the  pulse obtains as a result of self-evolution.

One important concept that arises from this is that of the \emph{matched} spot size -- \ie one where the self-focusing caused by the plasma balances the natural diffraction of the laser pulse and stable propagation occurs.
In the blow-out regime (where the laser pulse expels practically all the electrons from inside the bubble), the transverse density profile of the bubble is approximately zero and flat inside the bubble, with very steep walls at the edges.
 The refractive index of an underdense plasma is $\eta \approx  1 - n_\mathrm{e}/(2n_\mathrm{c}) $,  so the refractive index profile of this bubble is  similar to that of a  single-mode optical fibre, \ie we have a `core' of a certain diameter (the bubble diameter) surrounded by `cladding' (the bubble sheath) of a lower refractive index. The guided mode in such an optical fibre has a transverse size that is approximately equal to the size of the core, so we expect that in a laser wakefield accelerator we will get stable propagation (\ie no spot size oscillations) when the transverse size of the laser pulse is approximately equal to that of the bubble.

Of course, one main difference between the bubble and an optical fibre is that the size of the bubble is determined by the properties of the laser pulse itself.
If the laser spot is too small, then it will drive a bubble that is larger than the laser spot.
This  over-sized bubble will support a larger mode and the laser pulse will expand, which in turn reduces the bubble size.
The radius of the bubble can be found by balancing the ponderomotive force of the laser pulse with the force due to the electric field of the bubble.
If the bubble is approximately the same size as the laser pulse then it will be matched.
The size of the matched spot can be found by balancing the forces on an electron at the edge of the bubble (\ie the laser's ponderomotive force and the force due to the electric field inside the bubble).
An expression for the matched spot size, $w_\mathrm{m}$ is
\begin{equation}
\label{matched1}
w_\mathrm{m} \simeq 2 \frac{ c}{\omega_\mathrm{p}} \sqrt{a_0}\ ,
\end{equation}
where the numerical factor of two was found through particle-in-cell simulations\cite{Lu2007e}.
This expression is particularly useful for finding the correct initial parameters of the accelerator, \ie for a given laser system, one should first determine the spot size at which the threshold $a_0 \simeq 3$ is reached; this expression can then be used to determine the correct operating plasma density.
However, when self-focusing  plays an important role, we need an expression for the matched spot size that depends not on the initial laser intensity but rather  the value of $a_0$ that it reaches after self-focusing.
We can use the fact that the ratio of the laser power, $P_\mathrm{L}$,  to the critical power for self-focusing, $P_\mathrm{c}$, can be written as
\begin{equation}
\frac{P_\mathrm{L}}{P_\mathrm{c}} = \frac{1}{32} \frac{\omega_p^2}{c^2}\,{a_0^2 w^2}\ ,
\end{equation}
and the fact that $a_0^2 w^2$ is constant during focusing (assuming that self-focusing happens more rapidly than any pulse compression) to eliminate $a_0$ from \Eref{matched1}.
This results in the following expression for the matched spot size:
\begin{equation}
\label{matched2}
w_\mathrm{m} \simeq   2\sqrt{2}  \frac{c}{\omega_\mathrm{p}} \left(\frac{P_\mathrm{L}}{P_\mathrm{c}}\right)^{\frac{1}{6}}\ .
\end{equation}
Note that $2 w_\mathrm{m} \approx  \lambda_\mathrm{p}$, as long as $P_\mathrm{L}$ is not many times greater than $P_\mathrm{c}$.

\begin{figure}
\begin{center}
\includegraphics[width=10cm]{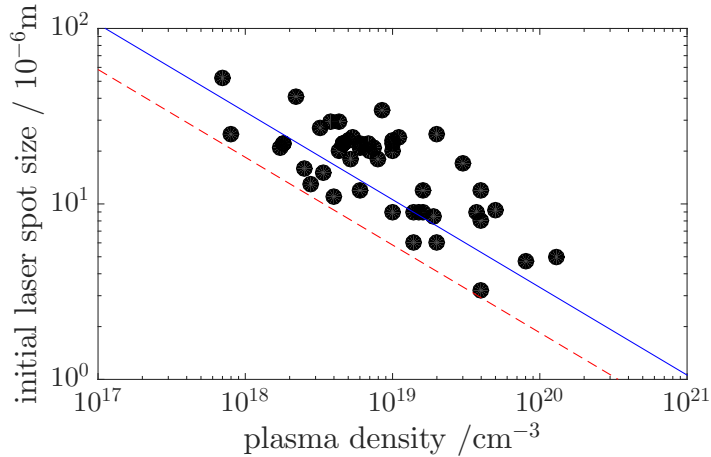}
\caption{Variation of initial laser focal spot size with operating plasma density in various laser wakefield acceleration experiments. Solid blue line, plasma wavelength, $\lambda_\mathrm{p}$; red dashed line, matched spot size assuming $a_0 \approx 3$.
Data from Refs.
\cite{Mangles2004,
 Geddes2004,
 Faure2004,
 Miura2005,
 Kotaki2006,
 Mori2006,
 Masuda2006,
 Hsieh2006,
 Hidding2006,
 Hosokai2006a,
 Mangles2006a,
 Faure2006,
 Reed2006,
 Leemans2006,
 Masuda2007,
 Ohkubo2007,
 Karsch2007,
 Mangles2007,
 Gamucci2008,
 Rowlands-Rees2008,
 Hafz2008a,
 Schmid2009a,
 Kneip2009b,
 Froula2009,
 Schmid2010a,
 Pak2010a,
 Ibbotson2010a,
 Clayton2010,
 McGuffey2010,
 Lu2011,
 Fourmaux2011a,
 Liu2011a,
 Pollock2011,
 Lundh2011,
 Gonsalves2011,
 Brijesh2012,
 Mo2012,
 Kneip2012,
 Weingartner2012,
 Albert2013,
 Walker2013,
 Mo2013a,
 Corde2013b,
 Chen2013a,
 Kim2013,
 Wang2013a,
 Sarri2014,
 Powers2014,
 Leemans2014,
 Khrennikov2015,
 Schnell2015
}.}
\label{spot_size}
\end{center}
\end{figure}

It is interesting to examine how the spot size used in experiments compares with this matched spot size.
Figure \ref{spot_size} shows the variation in the initial (\ie vacuum) laser focal spot size with the operating plasma density in  the experiments in the dataset.
There is a clear overall trend towards larger initial spots at lower densities (and therefore higher electron beam energies), and the spot sizes used are reasonably close to $\lambda_\mathrm{p}$ (shown as a solid blue line in  \Fref{spot_size}).
However, given that the evidence suggests that most of these experiments reached $a_0 \approx 3$ as a result of pulse evolution, it is also interesting to compare the initial spot size with the expected matched spot size for $a_0 \approx 3$ (this is shown as a dashed red line in \Fref{spot_size}).
Most of the experiments are clearly operating at an initial spot size significantly larger than this matched spot size, and none of the selected experiments operate with a spot size below this.
This suggests  either that most experiments are operating with too large an initial spot size, wasting accelerator length and laser energy while the laser pulse self-focuses, or that there is some experimental advantage in starting at a spot size larger than the matched spot size and letting pulse evolution happen.

\subsection{To guide or not to guide?}
\begin{figure}
\begin{center}
\includegraphics[width=10cm]{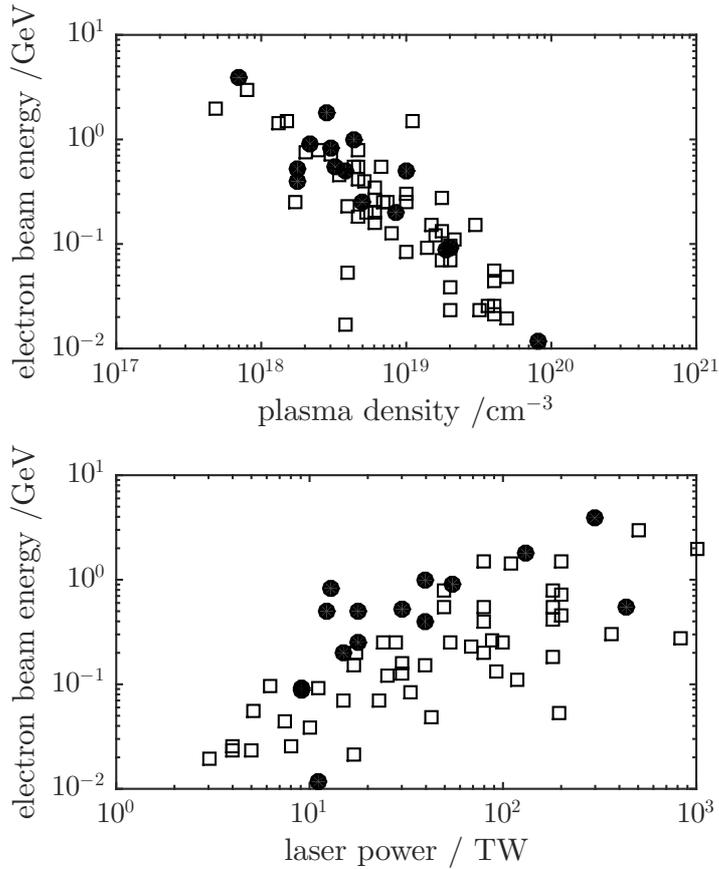}
\caption{Variation of reported electron beam energy from various experiments. Top: Variation of beam energy with plasma density. Bottom: Variation of beam energy with laser power. Filled circles, experiments in pre\-formed guiding structures; open squares, experiments without preformed guiding structures.
Data from Refs.
\cite{Mangles2004,
 Geddes2004,
 Faure2004,
 Miura2005,
 Kotaki2006,
 Mori2006,
 Masuda2006,
 Hsieh2006,
 Hidding2006,
 Hosokai2006a,
 Mangles2006a,
 Faure2006,
 Reed2006,
 Leemans2006,
 Masuda2007,
 Ohkubo2007,
 Karsch2007,
 Mangles2007,
 Gamucci2008,
 Rowlands-Rees2008,
 Hafz2008a,
 Schmid2009a,
 Kneip2009b,
 Froula2009,
 Schmid2010a,
 Pak2010a,
 Ibbotson2010a,
 Clayton2010,
 McGuffey2010,
 Lu2011,
 Fourmaux2011a,
 Liu2011a,
 Pollock2011,
 Lundh2011,
 Gonsalves2011,
 Brijesh2012,
 Mo2012,
 Kneip2012,
 Weingartner2012,
 Albert2013,
 Walker2013,
 Mo2013a,
 Corde2013b,
 Chen2013a,
 Kim2013,
 Wang2013a,
 Sarri2014,
 Powers2014,
 Leemans2014,
 Khrennikov2015,
 Schnell2015
}.}
\label{guiding_type}
\end{center}
\end{figure}

It is well known that a tightly focused laser pulse will quickly diffract in a vacuum.
Effectively, the pulse can only remain intense over a distance of about a Rayleigh length, $z_\mathrm{R} = \pi w_0^2/\lambda$.
The accelerator lengths used in laser wakefield experiments are typically much longer than this and some sort of `guiding' is therefore needed to keep the laser intensity sufficiently high to drive a wake throughout the structure.
There are two principal techniques to achieve this guiding, which both rely on creating a waveguide structure to counteract diffraction.
This requires that the transverse plasma density profile has a minimum on-axis.
Such a density minimum is naturally created by the laser pulse itself -- that is, the bubble itself acts as a waveguide.
Alternatively, a preformed density minimum can be formed, for example using a  capillary discharge \cite{Leemans2006}; using a preformed waveguide brings significant complexity to an experiment and restricts diagnostic access (\eg for  wake imaging diagnostics based on ultrashort probes \cite{Savert2015}).
However, the bubble cannot self-guide the very front slice of the laser pulse, since the density minimum is not formed immediately.
Because of this, external channels  are expected to be more efficient, \ie they allow the laser to propagate over greater distances at high $a_0$.
With these points in mind, it is interesting to see whether the experimental evidence supports the use of external waveguides.
In \Fref{guiding_type}, the data are sorted into self-guided and externally guided experiments.
The top panel of \Fref{guiding_type} appears to show that there is no real advantage to using external guiding structures; the electron beam energy, as a function of plasma density, follows the same trend for both subsets of the data, \ie they are both limited by dephasing.
However, the bottom panel of \Fref{guiding_type} reveals the distinct advantage that experiments in externally guided structures have over self-guided ones.
The highest electron energy achieved at a given laser power is almost always from an externally guided experiment; the self-guided experiments at the same power tend to produce lower energy electron beams.

\begin{figure}
\begin{center}
\includegraphics[width=10cm]{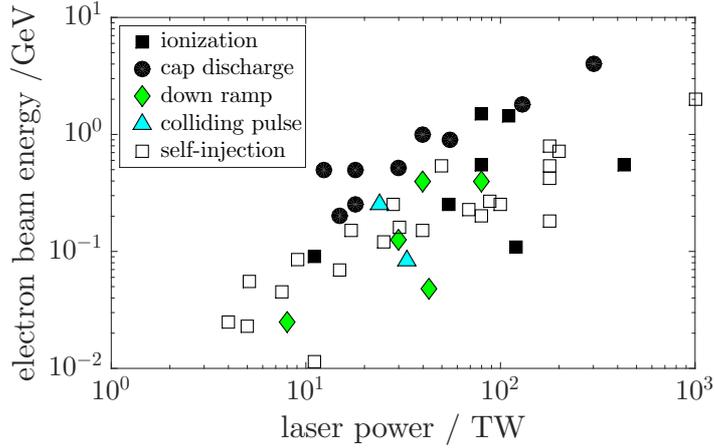}
\caption{Variation of reported electron beam energy from various experiments as a function of laser power and for different injection mechanisms.
Filled black circles, capillary discharge experiments;
black squares,  ionization injection;
green diamonds, density down-ramp injection;
cyan triangles, colliding pulse injection;
open black squares,  self-injection.
Data from Refs.
\cite{Mangles2004,
 Geddes2004,
 Faure2004,
 Miura2005,
 Kotaki2006,
 Mori2006,
 Masuda2006,
 Hsieh2006,
 Hidding2006,
 Hosokai2006a,
 Mangles2006a,
 Faure2006,
 Reed2006,
 Leemans2006,
 Masuda2007,
 Ohkubo2007,
 Karsch2007,
 Mangles2007,
 Gamucci2008,
 Rowlands-Rees2008,
 Hafz2008a,
 Schmid2009a,
 Kneip2009b,
 Froula2009,
 Schmid2010a,
 Pak2010a,
 Ibbotson2010a,
 Clayton2010,
 McGuffey2010,
 Lu2011,
 Fourmaux2011a,
 Liu2011a,
 Pollock2011,
 Lundh2011,
 Gonsalves2011,
 Brijesh2012,
 Mo2012,
 Kneip2012,
 Weingartner2012,
 Albert2013,
 Walker2013,
 Mo2013a,
 Corde2013b,
 Chen2013a,
 Kim2013,
 Wang2013a,
 Sarri2014,
 Powers2014,
 Leemans2014,
 Khrennikov2015,
 Schnell2015
}.}
\label{injection_type}
\end{center}
\end{figure}

Are these data sufficient to suggest that external guiding structures improve the efficiency of laser wakefield accelerators, owing to reduced energy losses as the pulse propagates?
Or is something more subtle occurring?
In one experiment with an external waveguide using a  ${\simeq}20\UW[T]$ laser \cite{Rowlands-Rees2008},  differ\-ences between  the plasma density measured during a high-intensity laser shot and plasma density measure\-ments made offline suggested that, under the conditions in which electron beams were observed, the laser pulse caused additional ionization of the plasma.
As the plasma used in that experiment was formed from hydrogen, which is readily ionized, this suggested that high-$Z$ impurities from the walls of the capillary were being ionized  by the main pulse and injected into the plasma wave.
Since that work, a number of groups have proceeded to exploit this ionization injection mechanism to reduce the threshold for injection and increase the beam charge (\eg Refs. \cite{Pak2010a,McGuffey2010}).
There are also various other injection techniques, including injection due to propagation in a density gradient \cite{Schmid2010a,Gonsalves2011} and colliding pulse in\-jection \cite{Faure2006}.
In \Fref{injection_type}, the variation of electron beam energy with laser power is plotted again, this time with the data divided into subsets based on injection type.
As it is not known whether the majority of capillary-discharge-based experiments rely on self-injection or whether ionization injection plays a role, as in Ref. \cite{Rowlands-Rees2008}, these experiments have been placed in their own subset. Some of the ionization in\-jection experiments also produce higher electron beam energies for a given laser power than self-injection experiments, and lie on the upper curve of the entire dataset, just as the capillary discharge dataset does. This might suggest that the injection mechanism  is more important than the guiding mechanism in determining the maximum energy that can be achieved from a given laser power.
This makes sense, as the value that $a_0$ reaches after  self-evolution decreases with decreasing laser power, but alternative injection mechanisms should lower the value that $a_0$  needs to reach in order for injection to occur.  However,  injection mechanisms other than ionization injection seem to perform similarly to self-injection in terms of the electron energy that can be obtained for a given laser power. At present, the evidence is still inconclusive; this is clearly a matter that requires further study.

\section{Future directions}

Having examined the trends in experimental data from laser wakefield experiments over the last decade, we find that there is one overriding message.
To keep pushing the electron energy achievable from a single stage of a laser wakefield to ever higher values clearly requires operation at lower densities and over longer distances, and such experiments will need more powerful lasers.
However, it should be noted that laser power can be increased in two ways, either by increasing the laser energy or by decreasing the pulse duration.
Most of the data presented were obtained using laser pulses with a duration of ${\sim}30\Ufs$, so the trends in laser power are dominated by the pulse energy.
Simulations demonstrate that the pulse length in a laser wakefield accelerator should not be too short.
Energy is lost  predominantly from the front of the laser pulse (this is the part of the laser that is doing work on the plasma),  so the shorter the laser pulse, the more quickly it runs out of energy.
As new laser facilities come online around the world, it is worth remembering that the shortest pulse is not necessarily the best for driving laser wakefield accelerators.
Pulse durations where $c\tau_\mathrm{L}$ is comparable to the plasma wavelength, $\lambda_\mathrm{p}$, drive efficient wakes and allow the pump depletion length to match the dephasing length \cite{Lu2007e}.
 Pulses significantly shorter than $\lambda_\mathrm{p}/2$ expend their energy before the maximum energy can be reached.

The experimental trends show that most experiments are attaining electron beam energies that are limited by dephasing.
If the community is to be able to keep up the impressive pace it has sustained over the last decade then methods to overcome this limitation will become increasingly important, especially as further increases in laser power become ever more expensive.
As a result, techniques to overcome dephasing, such as quasi-phase matching \cite{Yoon2012}, staging \cite{Kaganovich2005} and density tapering \cite{Sprangle2001} will all become important areas of research.

The final point that we would like to make is that the laser systems currently used to drive laser wakefield accelerators are woefully inefficient.
A titanium sapphire laser has a `wall plug' efficiency of ${\sim}0.1\%$.
They also currently run at very low repetition rates compared with a conventional accelerator.
As laser wakefield accelerators push the energy frontier and strive to become workhorses for applications, it will become increasingly important that both repetition rate and efficiency are properly considered and significantly improved.
Much more efficient laser architectures are available, including thin-disc  \cite{Giesen2007}  and fibre \cite{Jauregui2013} lasers, that can also readily operate at much higher repetition rates.  However, these sys\-tems do not yet have the capability to produce laser pulses with the high energy and short pulse duration needed to drive current laser wakefield acceleration experiments.
Innovative solutions, including the use of coherent \cite{Mourou2013} and incoherent \cite{Benedetti2014} combinations of many low-energy laser pulses to drive  wake\-field experiments, or the use of trains of many low-energy pulses to resonantly drive a high amplitude wakefield \cite{Hooker2014}, may well form the  most promising routes to high repetition rate, high-efficiency laser wakefield accelerators suitable for particle physics experiments or light-source based applications.



\begin{thebibliography}{10}
\raggedright
\bibitem{Tajima1979}
T.~Tajima and J.M. Dawson, \newblock \textit{Phys. Rev. Lett.} {\bf 43}(4) (1979) 267. http://dx.doi.org/10.1103/PhysRevLett.43.267

\bibitem{Strickland1985}
D.~Strickland and G.~Mourou, \newblock \textit{Opt. Commun.} {\bf 55}(6) (1985), 447--449. http://dx.doi.org/10.1016/0030-4018(85)90151-8

\bibitem{Nakajima1994}
K.~Nakajima {\em et~al.}, \newblock \textit{Phys. Scr.} {\bf T52} (1994) 61. http://dx.doi.org/10.1088/0031-8949/1994/T52/009

\bibitem{Amiranoff1998}
F.~Amiranoff {\em et~al.}, \newblock \textit{Phys. Rev. Lett.} {\bf 81}(5) (1998) 995. http://dx.doi.org/10.1103/PhysRevLett.81.995

\bibitem{Mangles2004}
S.P.D. Mangles {\em et~al.}, \newblock \textit{Nature} {\bf 431}
(2004) 535--538. http://dx.doi.org/10.1038/nature02939

\bibitem{Geddes2004}
C.G.R. Geddes {\em et~al.}, \newblock
\textit{Nature} {\bf 431} (2004)  538--541. http://dx.doi.org/10.1038/nature02900

\bibitem{Faure2004}
J.~Faure {\em et~al.}, \newblock \textit{Nature} {\bf 431} (2004) 541--544. http://dx.doi.org/10.1038/nature02963

\bibitem{Faure2006}
J.~Faure {\em et~al.}, \newblock \textit{Nature} {\bf 444} (2006)
737--739. http://dx.doi.org/10.1038/nature05393

\bibitem{Gonsalves2011}
A.J. Gonsalves {\em et~al.}, \newblock \textit{Nat. Phys.} {\bf
7} (2011) 862--866. http://dx.doi.org/10.1038/nphys2071

\bibitem{Matlis2006}
N.H. Matlis {\em et~al.}, \newblock
\textit{Nat. Phys.} {\bf 2} (2006) 749--753. http://dx.doi.org/10.1038/nphys442

\bibitem{Savert2015}
A.~S\"avert {\em et~al.},\newblock \textit{Phys. Rev. Lett.} {\bf 115}(5) (2015) 055002. http://dx.doi.org/10.1103/PhysRevLett.115.055002


\bibitem{Rousse2004}
A.~Rousse {\em et~al.}, \newblock \textit{Phys. Rev. Lett.} {\bf 93}(13) (2004) 135005. http://dx.doi.org/10.1103/PhysRevLett.93.135005

\bibitem{Kneip2010}
S.~Kneip {\em et~al.}, \newblock \textit{Nat. Phys.} {\bf 6}
(2010) 980--983. http://dx.doi.org/10.1038/nphys1789

\bibitem{Fuchs2009b}
M.~Fuchs {\em et~al.}, \newblock \textit{Nat. Phys.} {\bf 5}
(2009) 826--829. http://dx.doi.org/10.1038/nphys1404

\bibitem{Miura2005}
E.~Miura {\em et~al.}, \newblock \textit{Appl. Phys. Lett.} {\bf
86}(25) (2005) 251501. http://dx.doi.org/10.1063/1.1949289

\bibitem{Kotaki2006}
H.~Kotaki {\em et~al.}, \newblock \textit{Laser Phys.} {\bf 16}(7)
(2006) 1107--1110. http://dx.doi.org/10.1134/S1054660X06070140

\bibitem{Mori2006}
M.~Mori {\em et~al.}, \newblock \textit{Phys. Lett.} A {\bf
356}(2) (2006) 146--151. http://dx.doi.org/10.1016/j.physleta.2006.06.001

\bibitem{Masuda2006}
S.~Masuda {\em et~al.}, \newblock \textit{J. Phys. IV France}
{\bf 133} (2006) 1127--1129.  http://dx.doi.org/10.1051/jp4:2006133229

\bibitem{Hsieh2006}
C.-T. Hsieh {\em et~al.}, \newblock \textit{Phys. Rev. Lett.}
{\bf 96}(9) (2006) 095001. http://dx.doi.org/10.1103/PhysRevLett.96.095001

\bibitem{Hidding2006}
B.~Hidding {\em et~al.}, \newblock \textit{Phys. Rev. Lett.} {\bf
96}(10) (2006) 105004. http://dx.doi.org/10.1103/PhysRevLett.96.105004

\bibitem{Hosokai2006a}
T.~Hosokai {\em et~al.}, \newblock \textit{Phys. Rev.} E {\bf
73}(3)
(2006) 036407. http://dx.doi.org/10.1103/PhysRevE.73.036407

\bibitem{Mangles2006a}
S.P.D. Mangles {\em et~al.}, \newblock \textit{Phys. Rev. Lett.}
{\bf 96}(21) (2006) 215001. http://dx.doi.org/10.1103/PhysRevLett.96.215001

\bibitem{Reed2006}
S.A. Reed {\em et~al.}, \newblock \textit{Appl. Phys. Lett.} {\bf
89}(23) (2006) 231107. http://dx.doi.org/10.1063/1.2400400

\bibitem{Leemans2006}
W.P. Leemans {\em et~al.}, \newblock \textit{Nat. Phys.} {\bf 2}
(2006) 696--699. http://dx.doi.org/10.1038/nphys418

\bibitem{Masuda2007}
S.~Masuda {\em et~al.}, \newblock \textit{Phys. Plasmas} {\bf 14}(2)
(2007) 023103.  http://dx.doi.org/10.1063/1.2434248

\bibitem{Ohkubo2007}
T.~Ohkubo {\em et~al.}, \newblock \textit{Phys. Rev. ST Accel.
Beams} {\bf 10}(3) (2007) 031301. http://dx.doi.org/10.1103/PhysRevSTAB.10.031301

\bibitem{Karsch2007}
S.~Karsch {\em et~al.}, \newblock \textit{New J. Phys.} {\bf 9}
(2007) 415. http://dx.doi.org/10.1088/1367-2630/9/11/415

\bibitem{Mangles2007}
S.P.D.~Mangles {\em et~al.}, \newblock \textit{Phys. Plasmas}
{\bf 14}(5) (2007) 056702.  http://dx.doi.org/10.1063/1.2436481

\bibitem{Gamucci2008}
A.~Gamucci {\em et~al.}, \newblock \textit{IEEE Trans. Plasma
Sci.} {\bf 36}(4) (2008) 1699--1706. http://dx.doi.org/10.1109/TPS.2008.2000898

\bibitem{Rowlands-Rees2008}
T.P. Rowlands-Rees {\em et~al.}, \newblock \textit{Phys. Rev.
Lett.} {\bf 100}(10) (2008) 105005. http://dx.doi.org/10.1103/PhysRevLett.100.105005

\bibitem{Hafz2008a}
N.A.M. Hafz {\em et~al.}, \newblock \textit{Nat. Photonics} {\bf
2} (2008) 571. http://dx.doi.org/10.1038/nphoton.2008.155

\bibitem{Schmid2009a}
K.~Schmid {\em et~al.}, \newblock \textit{Phys. Rev. Lett.} {\bf
102}(12) (2009) 124801. http://dx.doi.org/10.1103/PhysRevLett.102.124801

\bibitem{Kneip2009b}
S.~Kneip {\em et~al.}, \newblock \textit{Phys. Rev. Lett.} {\bf
103}(3) (2009) 035002. http://dx.doi.org/10.1103/PhysRevLett.103.035002

\bibitem{Froula2009}
D.H. Froula {\em et~al.}, \newblock \textit{Phys. Rev. Lett.}
{\bf 103}(21) (2009) 215006. http://dx.doi.org/10.1103/PhysRevLett.103.215006

\bibitem{Schmid2010a}
K.~Schmid {\em et~al.}, \newblock \textit{Phys. Rev. ST Accel.
Beams} {\bf 13}(9) (2010) 091301. http://dx.doi.org/10.1103/PhysRevSTAB.13.091301

\bibitem{Pak2010a}
A.E. Pak {\em et~al.}, \newblock \textit{Phys. Rev.
Lett.} {\bf 104}(2) (2010) 025003. http://dx.doi.org/10.1103/PhysRevLett.104.025003

\bibitem{Ibbotson2010a}
T.P.A. Ibbotson {\em et~al.}, \newblock \textit{New J. Phys.} {\bf 12} (2010) 45008. http://dx.doi.org/10.1088/1367-2630/12/4/045008

\bibitem{Clayton2010}
C.E. Clayton {\em et~al.}, \newblock \textit{Phys. Rev. Lett.}
{\bf 105}(10) (2010) 105003. http://dx.doi.org/10.1103/PhysRevLett.105.105003

\bibitem{McGuffey2010}
C.~McGuffey {\em et~al.}, \newblock \textit{Phys. Rev. Lett.} {\bf
104}(2) (2010) 025004. http://dx.doi.org/10.1103/PhysRevLett.104.025004

\bibitem{Lu2011}
H.~Lu {\em et~al.}, \newblock \textit{Appl. Phys. Lett.} {\bf 99}(9)
(2011) 091502. http://dx.doi.org/10.1063/1.3626042

\bibitem{Fourmaux2011a}
S.~Fourmaux {\em et~al.}, \newblock \textit{New J. Phys.} {\bf
13} (2011) 033017. http://dx.doi.org/10.1088/1367-2630/13/3/033017

\bibitem{Liu2011a}
J.S. Liu {\em et~al.}, \newblock
\textit{Phys. Rev. Lett.} {\bf 107}(3) (2011) 035001. http://dx.doi.org/10.1103/PhysRevLett.107.035001

\bibitem{Pollock2011}
B.B. Pollock {\em et~al.}, \newblock \textit{Phys. Rev. Lett.}
{\bf 107}(4) (2011) 045001. http://dx.doi.org/10.1103/PhysRevLett.107.045001

\bibitem{Lundh2011}
O.~Lundh {\em et~al.}, \newblock \textit{Nat. Phys.} {\bf 7}
(2011) 219--222. http://dx.doi.org/10.1038/nphys1872

\bibitem{Brijesh2012}
P.~Brijesh {\em et~al.}, \newblock \textit{Phys. Plasmas} {\bf
19}(6) (2012) 063104. http://dx.doi.org/10.1063/1.4725421

\bibitem{Mo2012}
M.Z. Mo {\em et~al.}, \newblock \textit{Appl. Phys. Lett.} {\bf
100}(7) (2012) 074101.  http://dx.doi.org/10.1063/1.3685464

\bibitem{Kneip2012}
S.~Kneip {\em et~al.}, \newblock \textit{Phys. Rev. ST Accel.
Beams} {\bf 15}(2) (2012)  021302. http://dx.doi.org/10.1103/PhysRevSTAB.15.021302

\bibitem{Weingartner2012}
R.~Weingartner {\em et~al.}, \newblock \textit{Phys. Rev. ST
Accel. Beams} {\bf 15}(11) (2012) 111302. http://dx.doi.org/10.1103/PhysRevSTAB.15.111302

\bibitem{Albert2013}
F.~Albert {\em et~al.}, \newblock \textit{Phys. Rev. Lett.} {\bf
111}(23) (2013) 235004. http://dx.doi.org/10.1103/PhysRevLett.111.235004

\bibitem{Walker2013}
P.A. Walker {\em et~al.}, \newblock \textit{New J. Phys.} {\bf
15} (2013) 045024. http://dx.doi.org/10.1088/1367-2630/15/4/045024

\bibitem{Mo2013a}
M.Z. Mo {\em et~al.}, \newblock \textit{Appl. Phys. Lett.} {\bf
102}(13) (2013) 134102. http://dx.doi.org/10.1063/1.4799280

\bibitem{Corde2013b}
S.~Corde {\em et~al.}, \newblock \textit{Nat. Commun.} {\bf 4} (1501) (2013) 1309.6364v1. http://dx.doi.org/10.1038/ncomms2528

\bibitem{Chen2013a}
S.~Chen {\em et~al.}, \newblock
\textit{Phys. Rev. Lett.} {\bf 110}(15) (2013) 155003. http://dx.doi.org/10.1103/PhysRevLett.110.155003

\bibitem{Kim2013}
H.T. Kim {\em et~al.}, \newblock \textit{Phys. Rev. Lett.} {\bf
111}(16) (2013) 165002. http://dx.doi.org/10.1103/PhysRevLett.111.165002

\bibitem{Wang2013a}
X.~Wang {\em et~al.}, \newblock \textit{Nat. Commun.} {\bf 4}(1988)
(2013). http://dx.doi.org/10.1038/ncomms2988

\bibitem{Sarri2014}
G.~Sarri {\em et~al.}, \newblock \textit{Phys. Rev. Lett.} {\bf
113} (2014) 224801. http://dx.doi.org/10.1103/PhysRevLett.113.224801

\bibitem{Powers2014}
N.D. Powers {\em et~al.}, \newblock \textit{Nat. Photonics} {\bf
8} (2014) 28--31. http://dx.doi.org/10.1038/nphoton.2013.314

\bibitem{Leemans2014}
W.P. Leemans {\em et~al.}, \newblock \textit{Phys. Rev. Lett.} {\bf
113}(24) (2014) 245002. http://dx.doi.org/10.1103/PhysRevLett.113.245002

\bibitem{Khrennikov2015}
K.~Khrennikov {\em et~al.}, \newblock \textit{Phys. Rev. Lett.} {\bf
114}(19) (2015) 195003. http://dx.doi.org/10.1103/PhysRevLett.114.195003

\bibitem{Schnell2015}
M.~Schnell {\em et~al.}, \newblock \textit{J. Plasma Phys.} {\bf
81}(04) (2015) 475810401.  http://dx.doi.org/10.1017/S0022377815000379

\bibitem{Malka2002}
V.~Malka {\em et~al.}, \newblock \textit{Science} {\bf 298} (5598)
(2002) 1596--1600. http://dx.doi.org/10.1126/science.1076782

\bibitem{Lu2007e}
W.~Lu {\em et~al.}, \newblock \textit{Phys. Rev. ST Accel. Beams}
{\bf 10}(6) (2007) 061301. http://dx.doi.org/10.1103/PhysRevSTAB.10.061301

\bibitem{Mori1997a}
W.B. Mori, \newblock \textit{IEEE J. Quant. Electron.} {\bf 33}(11)
(1997) 1942--1953. http://dx.doi.org/10.1109/3.641309

\bibitem{Yoon2012}
S.J. Yoon \textit{et al.}, \newblock \textit{Phys. Rev. ST Accel.
Beams} {\bf 15}(8) (2012) 081305. http://dx.doi.org/10.1103/PhysRevSTAB.15.081305

\bibitem{Kaganovich2005}
D.~Kaganovich {\em et~al.}, \newblock \textit{Phys. Plasmas} {\bf
12}(10) (2005) 100702. http://dx.doi.org/10.1063/1.2102727

\bibitem{Sprangle2001}
P.~Sprangle {\em et~al.}, \newblock \textit{Phys. Rev.} E {\bf
63}(5) (2001) 056405. http://dx.doi.org/10.1103/PhysRevE.63.056405

\bibitem{Giesen2007}
A.~Giesen and J.~Speiser, \newblock \textit{IEEE J. Sel. Top.
Quant. Electron.} {\bf 13}(3) (2007) 598. http://dx.doi.org/10.1109/JSTQE.2007.897180

\bibitem{Jauregui2013}
C.~Jauregui, J.~Limpert and A.~T\"unnermann,  \newblock \textit{Nat. Photonics} {\bf 7}
(2013) 861--867. http://dx.doi.org/10.1038/nphoton.2013.273

\bibitem{Mourou2013}
G.~Mourou \textit{et al.}, \newblock \textit{Nat. Photonics} {\bf 7}
(2013) 258--261. http://dx.doi.org/10.1038/nphoton.2013.75

\bibitem{Benedetti2014}
C.~Benedetti, \textit{et al.}, \newblock \textit{Phys. Plasmas} {\bf
21}(5) (2014)
056706. http://dx.doi.org/10.1063/1.4878620

\bibitem{Hooker2014}
S.M. Hooker {\em et~al.}, \newblock \textit{J. Phys. B: At. Mol. Opt.
Phys} {\bf 47}(23)
(2014) 234003. http://dx.doi.org/10.1088/0953-4075/47/23/234003
\end{thebibliography}

\end{document}